\documentclass[
 aps, pra,
 amsmath,amssymb,
 11pt,
 final,
tightenlines,
 twoside,
 twocolumn,
 nofloats,
nofootinbib,
 superscriptaddress,
showkeys,
showkeywords,
 ]
{revtex4}
\usepackage[utf8]{inputenc}
\usepackage[english]{babel}
\usepackage{graphicx}% Include figure files
\usepackage{dcolumn}% Align table columns on decimal point
\usepackage{bm}% bold math
\usepackage{longtable}

\input{maik.rty}

\setcitestyle{authoryear,round}
\def\saoname{Special Astrophysical Observatory,  Russian Academy of Sciences,
              Nizhnii Arkhyz, 369167 Russia}

\def\squareforqed{\hbox{\rlap{$\sqcap$}$\sqcup$}}

\def\sq{\ifmmode\squareforqed\else{\unskip\nobreak\hfil
\penalty50\hskip1em\null\nobreak\hfil\squareforqed
\parfillskip=0pt\finalhyphendemerits=0\endgraf}\fi}

\def\degr{\hbox{$^\circ$}}

\def\arcmin{\hbox{$^\prime$}}

\def\arcsec{\hbox{$^{\prime\prime}$}}

\def\utw{\smash{\rlap{\lower5pt\hbox{$\sim$}}}}

\def\udtw{\smash{\rlap{\lower6pt\hbox{$\approx$}}}}

\def\fm{\hbox{$\,.\!\!^{\rm m}$}}

\def\fdg{\hbox{$\,.\!\!^\circ$}}

\def\farcs{\hbox{$\,.\!\!^{\prime\prime}$}}

\def\diameter{{\ifmmode\mathchoice
{\ooalign{\hfil\hbox{$\displaystyle/$}\hfil\crcr
{\hbox{$\displaystyle\mathchar"20D$}}}}
{\ooalign{\hfil\hbox{$\textstyle/$}\hfil\crcr
{\hbox{$\textstyle\mathchar"20D$}}}}
{\ooalign{\hfil\hbox{$\scriptstyle/$}\hfil\crcr
{\hbox{$\scriptstyle\mathchar"20D$}}}}
{\ooalign{\hfil\hbox{$\scriptscriptstyle/$}\hfil\crcr
{\hbox{$\scriptscriptstyle\mathchar"20D$}}}}
\else{\ooalign{\hfil/\hfil\crcr\mathhexbox20D}}%
\fi}}

\newcommand{\ab}{Astrophysical Bulletin }
\newcommand{\aap}{Astron. and Astrophys. }
\newcommand{\aaps}{Astron. and Astrophys. Suppl. }
\newcommand{\aj}{Astron.~J. }
\renewcommand{\apj}{Astrophys.~J. }
\newcommand{\apjs}{Astrophys.~J. Suppl. }
\newcommand{\mnras}{Monthly Notices Royal Astron. Soc. }
\newcommand{\apjl}{Astrophys.~J.}

\def\be{\begin{equation}}
\def\ee{\end{equation}}
\def\ba{\begin{eqnarray}}
\def\ea{\end{eqnarray}}

\def\ltsima{$\; \buildrel < \over \sim \;$}
\def\simlt{\lower.5ex\hbox{\ltsima}}
\def\gtsima{$\; \buildrel > \over \sim \;$}
\def\simgt{\lower.5ex\hbox{\gtsima}}

\usepackage[dvipsnames]{color}
\begin{document}

\selectlanguage{english}
%\ydk{}
%\titlerunning{}
%\authorrunning{}
%\toctitle{}
\title{New Local Volume Dwarf Galaxy Candidates from the DESI
Legacy Imaging Surveys}
%\tocauthor{I.~D.~Karachentsev, E.~I.~Kaisina }

\author{\firstname{I.~D.}~\surname{Karachentsev}}
 \email{idkarach@gmail.com}
 \affiliation\saoname

\author{\firstname{E.~I.}~\surname{Kaisina}}
 %\email{ikar@sao.ru}
 \affiliation\saoname


%\received{ August 00,  2022} \revised{August 00,  2022} \accepted{August 00,  2022}

\begin{abstract}
We undertook a search for new dwarf galaxies in the vicinity of
relatively isolated nearby galaxies with distances $D < 12$~Mpc
and stellar masses in the ($2\times10^{11}$--$3\times10^8~M_{\odot}$)
interval, using the data from the DESI Legacy Imaging Surveys.
Around the 46~considered Local Volume galaxies, $67$ new candidates
for satellites of these galaxies were found. About half of
them are classified as spheroidal dwarfs of low surface brightness.
The new galaxies are included in the Local Volume database (LVGDB),
which now contains 1421~objects, being 63\% more than the Updated
Nearby Galaxy Catalog.

\end{abstract}

\maketitle

\section{INTRODUCTION}

Numerical modeling of the large scale structure of the Universe carried out by various international teams within the framework of the standard cosmological model $\Lambda$CDM \citep{kly1999,moo1999,tin2010,saw2016,kly2016} needs comparing with the observational data on galaxies that are contained in a fixed volume of space. However, most existing catalogs comprise samples limited by the apparent magnitude (flux) of the galaxies in one spectral region or another. The first attempt to create a catalog of galaxies located within a $10$~Mpc radius sphere around us was undertaken by \citet{kra1979}. That sample contained 179~galaxies with radial velocities $V_{\rm LG}<500$~km\,s$^{-1}$ relative to the Local Group (LG) centroid. Systematic searches for nearby dwarf galaxies over the   all  sky  \citep{kara1998,kar2000,kara1999} in the photographic prints of the Palomar survey (POSS-II, ESO-SERC) and subsequent measurements of their radial velocities \citep{huch2000,huch2001,huch2003} led to a significant  increase of the Local Volume (LV) galaxy sample. The Catalog of  Neighboring Galaxies (CNG)  contained 450~galaxies with expected  distances $D<10$~Mpc \citep{kar2004}.  Mass spectral surveys of large areas of the sky in the optical  region: 2dF \citep{col2003}, 6dF \citep{jon2009}, SDSS \citep{aba2009} and in the H\,I~21~cm radio line: HIPASS \citep{doy2005}; ALFALFA \citep{hay2018} have significantly enriched the Local Volume sample. Over nine years, the number of galaxies with $D<10$~Mpc has almost doubled. The Updated Nearby Galaxy Catalog (UNGC) now contained 869~objects \citep{kar2013}.

Various data on LV galaxies were collected in the
LVG database\footnote{\url{http://www.sao.ru/lv/lvgdb}}
\citep{kai2012}, which is regularly updated with new objects. To
date, this sample contains 1250~LV galaxy candidates with
\mbox{$D<11$}~Mpc distances. Evidently, the new additions are mainly due
to the dwarf galaxies with low integral luminosity and low surface
brightness.

We should note that a significant part of the galaxies in the LVG
database are merely candidates for this sample. Nearby galaxies
may have radial velocities that differ significantly from the
ideal Hubble velocity \mbox{$V_{\rm LG} = H_0D$}, where $H_0$ is
the Hubble parameter, due to the virial motions in nearby groups and
large scale flows. Thus, some LV galaxies take part in a
systematic flow towards the closest massive attractor in the Virgo
cluster as well as in the motions away from the center of the
expanding Local Void. The amplitude of both these flows reaches
about 200~km\,s$^{-1}$ \citep{tul2008}, which is comparable to
the quantity \mbox{$V_{\rm LG} <600$~km\,s$^{-1}$} used to include
galaxies in the LV.

Great progress in the LV sample formation was reached due to mass
distance measurements for galaxies based on the Hubble Space
Telescope (HST) data. Using the tip of the red giant branch (TRGB)
luminosity as a distance indicator allows one to determine the
distance to galaxies of any morphological type with an accuracy of
about 5\%. HST observations with the ACS camera in the ${F}814{W}$
and ${F}606{W}$ filters make it possible to measure the galaxy
distances by the TRGB method up to the LV boundary ($D\sim11$~Mpc)
over a single orbital period. A list of TRGB-distance estimates
for approximately 450~LV galaxies was compiled by \citet{ana2021}
and is available in the Extragalactic Distance Database
(EDD)\,\footnote{\url{http://edd.ifa.hawaii.edu}}.
  Obviously, high precision galaxy distance measurements lead both to an
  increase of the LV sample and to an exclusion of part of the objects from the nearby galaxy candidates.

In addition to the surveys of large areas of the northern and
southern sky, a noticeable contribution to the LV galaxy sample
additions is due to the searches for dwarf galaxies around nearby
massive Milky Way (MW) and M\,31 (Andromeda) type galaxies. The
entourage of galaxies accompanying the dominating Local Group
galaxies, MW and M\,31, already amounts to about~50~members each
\citep{kash2018,put2021}. \citet{chi2009,chi2013} carried out a
search for new dwarf galaxies in the vicinity of the neighboring
M\,81 galaxy using long-exposure frames obtained with the CFHT
telescope for subarcsecond images. These images enabled the
estimation of the TRGB position for dwarfs and thus to verify
their group membership. A similar approach was then used to search
for new satellites around NGC\,5128 (Cen\,A), NGC\,253, NGC\,628,
NGC\,4631, NGC\,4736  \citep{tan2017,grn2019,sme2018,
mul2019,dav2021,oka2019,tru2021,mut2022} using the CFHT and other
large telescopes (Subaru, VLT, GMT, LBT, Blanco) mounted in
locations with good astroclimate and equipped with wide-filed
CCD-cameras. Usually, the authors aimed to carry out surveys in
areas limited by the virial radius of the massive ``host'' galaxy.
For MW and  M\,31, the virial radius of the dark halo is
\mbox{$R_{\rm {vir}}\sim 250$~kpc}. CCD-images taken with
hours-long exposures using amateur small-aperture telescopes
\citep{jav2016,kar2020,mar2021} have also made a noticeable
contribution to the detection of nearby dwarf galaxy candidates.

The DESI~ Legacy~ Imaging~ Surveys \citep{dey2019}, covering wide
regions in the northern and southern sky, served as a significant
stimulus to search for dwarf galaxies of low surface brightness.
It is a combination of three projects: the Dark Energy Camera
Legacy Survey, the Beijing-Arizona Sky Survey, and the Mayall
z-band Legacy Survey. As of January 2021, this survey (DR\,9)
covers a sky area of about 14\,000~square degrees at galactic
latitudes $|b|>18\degr$.
Carried out in three optical bands, $g$, $r$ and $z$, this survey has a median $5\,\sigma$-limit of $r\simeq 23\fm1$ for galaxies with an exponential profile and effective radius $0\farcs5$, which is approximately $1^{\rm m}$ %одну звезднуі величину
deeper than the SDSS survey.
 \citet{car2022} used the {DESI Legacy Imaging Surveys} in combination with archival images obtained with the CFHT to search for satellites around massive MW-type LV galaxies. Besides the already known satellites, the authors found
68~new LV candidate members. They used the surface brightness
fluctuation method (sbf) (the main contribution to which is due to
the individually unresolvable red giant branch stars) to obtain
new distance estimates for 130~dwarf galaxies. This outstanding
paper also contains $g$- and
$r$-band photometry results for over 300~Local Volume galaxies.
We included the new data from \citet{car2022} in the LVG database. %\footnote{\url{http://www.sao.ru/lv/lvgdb}}.

\section{SEARCH FOR NEW DWARF GALAXIES}

The list of detected satellite candidates for 25~near\-by massive
galaxies \citep{car2022} has about 400~objects. More than 80\% of
them were already included in our database of LV galaxies. The
total area of the virial zones where \citet{car2022} searched for
new satellites amounts to 280~square degrees or 2\% of the total
area covered by the DESI Legacy Imaging Surveys. Evidently, the
remaining large area may contain other new candidate LV members.
An inspection of the entire remaining region of the { DESI Legacy
Imaging Surveys} is a rather difficult task. We therefore
restricted our search to dwarf satellites around the relatively
isolated nearby galaxies, less massive than the MW. We used the
list of LV galaxies with known satellites \citep{kar2021} as a
reference point.

The list of ``host'' galaxies around which we searched for new
satellites includes 46~galaxies and is presented in
Table~\ref{Karachentsev_table1}. The following quantities are
given in its columns: (1)---name of the galaxy; (2)---equatorial coordinates in degrees for the {J}\,2000.0 epoch; (3)
and (4)---distance to the galaxy in~Mpc and the method used to
determine it: TRGB---the tip of the red giant branch, SN---supernovae luminosity, sbf---surface brightness fluctuations,
TF---Tully-Fisher relation; NAM---radial velocity with
allowance for the peculiar velocity field in the Numerical Action
Method model, \citet{sha2017};
(5)---logarithm of the $K_s$-band
luminosity of the galaxy; (6) and (7)---angular and linear
projected distance that limited the search area for the galaxy
satellite candidates. Data on the galaxy distances and $K$-band
luminosities were taken from our updated LV galaxy database,
LVGDB, which also contains the data source references.

%\renewcommand{\baselinestretch}{0.9}
%\setcaptionmargin{0mm} \onelinecaptionstrue \captionstyle{normal}
   \begin{longtable*}{l|c|r@{~~~}|c|r@{~~~~}|c|c}
  \caption{List of survey regions around central galaxies
 \label{Karachentsev_table1}}\\ \hline

\hline
   Galaxy  &    RA (J\,2000)   {Dec},  deg &    $D$,  Mpc   &  Method   & $\log L_K$, $L_{\odot}$\!\!\!\!\!  &  $r_p$, deg  & $R_p$,  kpc\\

\hline
    \multicolumn{1}{c|}{(1)}    &    (2)       & \multicolumn{1}{c|}{(3) }  &   \multicolumn{1}{c|}{(4)}   &  \multicolumn{1}{c|}{(5)}    &   \multicolumn{1}{c|}{(6)}    &  \multicolumn{1}{c}{(7)}     \\
\hline
\endfirsthead
\caption{(Continued) } \\
\hline
   Galaxy  &    RA (J\,2000)  {Dec},  deg &    $D$,  Mpc   &  Method   & $\log L_K$, $L_{\odot}$\!\!\!\!\!  &  $r_p$, deg  & $R_p$,  kpc \\
\hline
     \multicolumn{1}{c|}{(1)}    &    (2)       & \multicolumn{1}{c|}{(3) }  &   \multicolumn{1}{c|}{(4)}   &  \multicolumn{1}{c|}{(5)}    &   \multicolumn{1}{c|}{(6)}    &  \multicolumn{1}{c}{(7)}     \\
\hline
\endhead

\hline
\endfoot

\hline
\endlastfoot
\hline
NGC\,628    &  ~ 24.174~~ $+$15.783 &    10.19&   TRGB  & 10.60    & 2.3    & 410    \\
NGC\,672    &  ~ 26.977~~ $+$27.433 &     7.18&   TRGB  &    9.65  &   2.0  &  250    \\
UGC\,1281   &  ~ 27.382~~ $+$32.588 &     5.27&   TRGB  &    8.57  &   0.5  &   50  \\
NGC\,784    &  ~ 30.321~~ $+$28.837 &     5.37&   TRGB  &    8.67  &   1.6  &  150    \\
NGC\,855    &  ~ 33.515~~ $+$27.877 &     9.73&   sbf   &    9.37  &   1.2  &  205    \\
NGC\,1744   &  ~ 74.991~~ $-$26.022 &    10.00&   TF    &    9.42  &   1.1  &  190   \\
NGC\,2337   &  107.556~~ $+$44.457 &    11.86&   TRGB  &    9.34  &   0.8  &  165    \\
NGC\,2366   &  112.223~~ $+$69.212 &     3.28&   TRGB  &    8.70  &   1.1  &   65   \\
NGC\,2403   &  114.202~~ $+$65.603 &     3.19&   TRGB  &    9.86  &   2.7  &  150   \\
NGC\,2683   &  133.172~~ $+$33.421 &     9.82&   TRGB  &   10.81  &   1.8  &  310   \\
NGC\,2787   &  139.828~~ $+$69.203 &     7.48&   sbf   &   10.19  &   2.1  &~~\,280* \\
NGC\,2903   &  143.042~~ $+$21.502 &     9.17&   TRGB  &   10.85  &   1.9  &  305   \\
M\,81       &  148.888~~ $+$69.065 &     3.70&   TRGB  &   10.95  &   6.2  &  400   \\
NGC\,3184   &  154.570~~ $+$41.424 &    11.12&   SN    &   10.52  &   2.3  &  450   \\
NGC\,3239   &  156.276~~ $+$17.162 &    10.17&   SN    &    9.74  &   1.2  &  210  \\
NGC\,3344   &  160.880~~ $+$24.922 &     9.82&   TRGB  &   10.33  &   1.5  &  255  \\
NGC\,3432   &  163.130~~ $+$36.619 &     9.14&   TRGB  &    9.64  &   1.4  &  220 \\
NGC\,3556   &  167.879~~ $+$55.674 &     9.90&   TF    &   10.52  &   1.5  &  260   \\
NGC\,3627   &  170.062~~ $+$12.992 &    11.12&   TRGB  &   11.08  &   3.0  &  570   \\
NGC\,3990   &  179.398~~ $+$55.459 &    10.30&   sbf   &    9.52  &   1.1  &  ~~\,200* \\
NGC\,4136   &  182.324~~ $+$29.927 &     6.67&   NAM   &    9.24  &   1.3  &  ~~\,150*\\
NGC\,4144   &  182.494~~ $+$46.457 &     6.89&   TRGB  &    9.25  &   1.3  &  155 \\
NGC\,4204   &  183.808~~ $+$20.659 &     7.01&   NAM   &    9.12  &   1.1  &  140 \\
NGC\,4236   &  184.176~~ $+$69.463 &     4.41&   TRGB  &    9.61  &   2.1  &  160 \\
NGC\,4242   &  184.376~~ $+$45.619 &     7.62&   TRGB  &    9.47  &   1.1  &  150 \\
NGC\,4244   &  184.374~~ $+$37.807 &     4.31&   TRGB  &    9.52  &   2.0  &  150\\
NGC\,4395   &  186.454~~ $+$33.547 &     4.76&   TRGB  &    9.47  &   1.8  &  150 \\
NGC\,4449   &  187.048~~ $+$44.090 &     4.27&   TRGB  &    9.68  &   2.0  &  150 \\
NGC\,4460   &  187.190~~ $+$44.864 &     7.28&   NAM   &    9.66  &   1.2  &  155\\
NGC\,4490   &  187.651~~ $+$41.644 &     8.91&   TRGB  &   10.28  &   2.0  &  310 \\
NGC\,4517   &  188.190~~ $+$00.115 &     8.36&   TRGB  &   10.27  &   1.7  &  250  \\
NGC\,4559   &  188.990~~ $+$27.960 &     8.91&   TRGB  &   10.20  &   1.7  &  265 \\
NGC\,4592   &  189.827~~ $-$00.532 &     9.08&   TRGB  &    9.16  &   0.7  &  110 \\
NGC\,4594   &  189.998~~ $-$11.623 &     9.55&   TRGB  &   11.32  &   3.2  &  ~~\,525* \\
NGC\,4597   &  190.054~~ $-$05.797 &    10.10&   TF    &    9.48  &   0.6  &  105\\
NGC\,4600   &  190.096~~ $+$03.118 &     9.29&   TRGB  &    9.33  &   0.6  &  100\\
NGC\,4605   &  189.997~~ $+$61.609 &     5.55&   TRGB  &    9.70  &   2.1  &  205 \\
NGC\,4618   &  190.387~~ $+$41.151 &     7.66&   TRGB  &    9.62  &   1.5  &  200 \\
NGC\,4631   &  190.533~~ $+$32.542 &     7.35&   TRGB  &   10.49  &   2.4  &  305 \\
NGC\,4736   &  192.721~~ $+$41.120 &     4.41&   TRGB  &   10.56  &   5.3  &  410 \\
NGC\,4861   &  194.760~~ $+$34.859 &     9.95&   TRGB  &    9.13  &   1.2  &  205 \\
NGC\,5023   &  198.052~~ $+$44.041 &     6.05&   TRGB  &    9.01  &   1.2  &  125 \\
NGC\,5055   &  198.954~~ $+$42.030 &     9.04&   TRGB  &   11.00  &   2.9  &  455 \\
NGC\,5194   &  202.469~~ $+$47.195 &     8.40&   SN    &   10.97  &   3.1  &  450 \\
NGC\,6503   &  267.360~~ $+$70.144 &     6.25&   TRGB  &   10.00  &   1.8  &  200 \\
IC\,5201    &  335.239~~ $-$46.036 &    10.40&   TF    &   10.15  &   1.3  &  235 \\
\hline
%\end{tabular}
\end{longtable*}

  According to \citet{tul2015}, the virial radius $R_{\rm {vir}}$ and total mass of the galaxy halo $M_T$ are connected by an empirical relation
    $$ (R_{\rm {vir}}/215\, {\rm kpc}) = (M_T/10^{12}M_{\odot})^{1/3}.$$

The total galaxy $K$-band luminosity is often used to estimate
$M_T$, assuming
\mbox{$M_T\!/\!L_K\!\!\sim\!\!(20$--$30)M_{\odot}/L_{\odot}$.} However, this
relation varies noticeably as a function of stellar mass, as well
as the galaxy morphological type \citep{kou2017,kar2022}. The
choice of $R_{\rm {vir}}$  and $R_p$ is therefore rather
subjective.

The surroundings of 11~massive MW type galaxies presented in
Table~\ref{Karachentsev_table1} were investigated by
\citet{car2022}. We have repeated the search for dwarf galaxies of
low surface brightness with the survey zone expanded beyond the
virial radius. Strictly speaking, a search for satellites should
be carried out within the limits of the zero velocity sphere
radius $R_0$, where the gravitational influence of the ``host''
galaxy dominates over the general cosmological expansion. However,
for a typical ratio \mbox{$R_0/R_{\rm {vir}}\simeq 3.5$}, the
searching task becomes ten times harder.

  Four galaxies in Table~\ref{Karachentsev_table1} are marked by an asterisk. For some reason or other, the distance estimates obtained for them may not match the real values. The distance estimate ($7.48$~Mpc) for NGC\,2787 made by surface brightness fluctuations is not reliable. There are also indirect indications that this galaxy may be located at \mbox{$D\sim 12$~Mpc.} Another, more distant, galaxy NGC\,3998 near
NGC\,3990 is located at $14.2$~Mpc (sbf). The retinues of both
galaxies are practically indistinguishable from each other. The
NGC\,4136 galaxy is located in a region of a specific scattered
group Coma\,I around NGC\,4150, where galaxies with velocities
\mbox{$V_{\rm LG}\sim (100$--$300)$~km\,s$^{-1}$} have distances
of about~$15$~Mpc. For the
NGC\,4594 (Sombrero) galaxy, the southern boarder of {DESI Legacy Imaging Surveys}
passes through the declination of \mbox{$-8\fdg5$,} and therefore our
search for satellites covers only a small part of this galaxy's
virial zone.

Note also that the results of our search for new dwarf~ galaxies~
around~ the~ neighbor~ galaxies NGC\,3115 and NGC\,3521 were reported
earlier \citep{kar2022}. The rich group around NGC\,3379 has a
virial radius noticeably larger than $300$~kpc, the value used to
search for dwarfs in \citet{car2022}. We suggest revisiting the
survey of this group's surroundings later.

Our searches for possible satellites around 46~LV galaxies led to
the discovery of 67~dwarf galaxies.
  %{ 50~ из которvх не представленv в базах даннvх LEDA \citep{mak2014} и
 % %NED \url{(http://ipac.caltech.edu)}.
 % NED\footnote{\url{(http://ipac.caltech.edu)}}.
 Their images taken from the DESI Legacy Imaging Surveys catalog are presented in the form of a mosaic in the Appendix. The size of each image corresponds to 120\arcsec, the north is at the top, the east is to the left.

The data on the new candidates in the LV are presented in the
columns of Table~\ref{Karachentsev_table2}:
  (1)---galaxy name;
  (2)---equatorial coordinates (J\,2000.0);
  (3)---maximally visible angular diameter of the galaxy in minutes of arc, approximately corresponding to the  Holmberg isophote;
(4)---apparent axial ratio; (5)---total $B$-band apparent
magnitude; for relatively bright galaxies, the $B$ values are
taken from LEDA and NED, and for fainter galaxies the apparent
magnitude was estimated by comparison with other galaxies of
similar structure with $g$- and $r$-magnitudes measured by
\citet{car2022}; the relation \mbox{$B=g+0.313(g-r)+0.227$} was
used to determine $B$, proposed by
% Lupton\footnote{\url{http://www.sdss3.org/dr10/algorithms/sdssUBVRITransform.php/}};
Lupton
(see~\url{http://www.sdss3.org/dr10/algorithms/sdssUBVRITransform.php/});
the accuracy of our visual estimations of $B$ amounts to about
$0\fm25$; (6)---morphological type of the galaxy determined by
us: Sph---spheroidal, Irr---irregular, Tr---transition
type between Sph and Irr, BCD---blue compact dwarf, dE---dwarf
elliptical galaxy; galaxies of high (H), normal (N), low (L) and
very low (VL) surface brightness are marked by the specified
letters;  {based on the photometry results for similar objects
\citep{car2022}, we estimated that the central $r$-band surface
brightness of our galaxies lies in the
\mbox{($22.5$--$26.5$)~mag/sq.arcsec} range;} (7)---distance to
the galaxy in Mpc; (8)---method used to estimate the distance:
mem---by the assumed membership in the entourage of the host
galaxy; TF or bTF---by the usual or baryonic Tully-Fisher
relation, NAM---by the radial velocity with allowance for
peculiar velocities due to local flows; sbf---by the surface
brightness fluctuations; txt---a rough distance estimate by the
texture of the galaxy.

%\renewcommand{\baselinestretch}{0.9}
%\setcaptionmargin{0mm} \onelinecaptionstrue \captionstyle{normal}
\begin{longtable*}{l|c|c|@{~~}c@{~~}|c|r@{}l|c|c}
\caption{New candidate LV galaxies from the {DESI Legacy Imaging Surveys}  \label{Karachentsev_table2}}\\
\hline
\multicolumn{1}{c|}{Name}        &      RA  (J2000)   Dec  &   $a_{\rm Ho}$,   arcmin   &   $b/a$   &  $B_T$ , mag &    \multicolumn{2}{c|}{Type}       &   $D$,   Mpc  &   Method  \\
\hline \multicolumn{1}{c|}{(1)}          &         (2)
&    (3)           &    (4)    &  (5)    &
\multicolumn{2}{c|}{(6)}  &   \multicolumn{1}{c|}{(7)}   &  (8)
\\ \hline
\endfirsthead
\caption{(Continued) }\\
\hline
\multicolumn{1}{c|}{Name}        &      RA  (J2000)  Dec   &   $a_{\rm Ho}$,   arcmin   &   $b/a$   &  $B_T$ , mag &    \multicolumn{2}{c|}{Type}       &   $D$,   Mpc  &   Method  \\
\hline \multicolumn{1}{c|}{(1)}          &         (2)
&    (3)           &    (4)    &  (5)    &
\multicolumn{2}{c|}{(6)}  &   \multicolumn{1}{c|}{(7)}   &  (8)
\\ \hline
\endhead
\hline
\endfoot
\hline
\endlastfoot
\hline
 Dw\,0143$+$1338 &    01 43 55.2~~ $+$13 38 42 &   0.86   &   0.86  & 18.6     & Sph{--}&VL  &    12.3  &   mem   \\
 Dw\,0149$+$3237 &    01 49 50.9~~ $+$32 37 42 &   0.40   &   0.95  & 21.5     & Sph{--}&VL  &     5.3  &   mem    \\
 Dw\,0158$+$3018 &    01 58 54.7~~ $+$30 18 50 &   0.67   &   0.40  & 18.4     & Tr{--}&L  &     5.4  &   mem    \\
 Dw\,0214$+$2836 &    02 14 09.6~~ $+$28 36 47 &   0.44   &   0.48  & 19.0     & Irr{--}&L  &     9.7  &   mem   \\
 Dw\,0218$+$2813 &    02 18 04.1~~ $+$28 13 01 &   0.72   &   0.91  & 21.3     & Sph{--}&VL  &     9.7  &   mem    \\
 $[$KKS\,2000$]$05&   02 49 26.1~~ $-$13 12 42 &   1.02   &   0.43  & 17.3       & Irr{--}&L  &    10.0  &   txt   \\
 Dw\,0827$+$6452 &    08 27 16.3~~ $+$64 52 26 &   0.45   &   0.87  & 20.5     & Sph{--}&L  &     3.7  &   mem    \\
 PGC\,025409     &    09 02 50.6~~ $+$71 18 22 &   1.16   &   0.73  & 16.3       & BCD{--}&N  &     7.5  &   mem    \\
 Dw\,0910$+$7326 &    09 10 15.6~~ $+$73 26 24 &   2.35   &   0.88  & 17.0     & Sph{--}&L  &     3.7  &   mem    \\
 Dw\,0910$+$6942 &    09 10 42.1~~ $+$69 42 11 &   0.37   &   0.82  & 19.1     & Sph{--}&L  &     7.5  &   mem    \\
 Dw\,0916$+$6944 &    09 16 43.7~~ $+$69 44 01 &   0.57   &   0.33  & 19.3     & Irr{--}&L  &     7.5  &   mem   \\
 Dw\,0918$+$6935 &    09 18 34.9~~ $+$69 35 43 &   0.27   &   0.85  & 20.0     & Sph{--}&L  &     7.5  &   mem   \\
 Dw\,0919$+$6932 &    09 19 07.8~~ $+$69 32 54 &   0.30   &    0.62 &  20.1    &  Tr{--}&L &     7.5 &    mem  \\
 UGC\,4918       &    09 19 17.7~~ $+$69 48 04 &    1.00  &    0.53 &  16.0      &  Im{--}&N &     7.5 &    mem  \\
 Dw\,0919$+$6955 &    09 19 33.9~~ $+$69 55 20 &    0.42  &    0.65 &  19.5     &  Sph{--}&L &     7.5 &    mem  \\
 Dw\,0920$+$6924 &    09 20 02.6~~ $+$69 24 45 &    0.18  &    0.80 &  20.4     &  Sph{--}&L &     7.5 &    mem  \\
 Dw\,0920$+$7017 &    09 20 22.4~~ $+$70 17 29 &    0.48  &    0.51 &  19.3     &  Sph{--}&L &     7.5 &    mem  \\
 Dw\,0927$+$6818 &    09 27 27.8~~ $+$68 18 55 &    0.53  &    0.76 &  19.6     &  Irr{--}&L &     7.5 &    mem   \\
 Dw\,1012$+$4259 &    10 12 42.7~~ $+$42 59 31 &    0.62  &    0.75 &  19.0     &  Tr{--}&N &    11.1 &    mem    \\
 Dw\,1051$+$6416 &    10 51 16.1~~ $+$64 16 41 &    0.33  &    0.93 &  21.0     &  Sph{--}&L &     9.0 &    txt   \\
 Dw\,1051$+$3617 &    10 51 27.6~~ $+$36 17 10 &    0.28  &    0.73 &  20.5     &  Sph{--}&L &     9.1 &    mem    \\
 KDG\,74         &    11 02 21.8~~ $+$70 15 50 &    0.85  &    0.67 &  18.6       &  Sph{--}&L &     3.7 &    mem   \\
 Dw\,1108$+$5520 &    11 08 59.5~~ $+$55 20 28 &    0.33  &    0.80 &  19.5     &  Sph{--}&L &     9.9 &    mem   \\
 Dw\,1109$+$5447 &    11 09 13.2~~ $+$54 47 10 &    0.34  &    0.78 &  19.7     &  Irr{--}&L &     9.9 &    mem    \\
 Dw\,1111$+$1338 &    11 11 35.0~~ $+$13 38 38 &    0.32  &    0.90 &  20.5     &  Sph{--}&L &    11.1 &    mem    \\
 Dw\,1113$+$5541 &    11 13 10.1~~ $+$55 41 17 &    0.42  &    0.64 &  20.5     &  Tr{--}&L &     9.9 &    mem   \\
 Dw\,1119$+$1011 &    11 19 30.7~~ $+$10 11 56 &    0.46  &    0.65 &  18.1     &  BCD{--}&H &    11.1 &    mem   \\
 Dw\,1127$+$1346 &    11 27 13.0~~ $+$13 46 52 &    0.35  &    0.92 &  20.1     &  Sph{--}&L &    11.1 &    mem   \\
 UGC\,6451       &    11 28 46.4~~ $+$79 36 07 &    2.66  &    0.53 &  16.5       &  Im{--}&N  &     3.7 &    mem   \\
 Dw\,1159$+$5554 &    11 59 56.6~~ $+$55 54 54 &    1.50  &    0.60 &  19.7     &  Tr{--}&L &     10.3 &    mem   \\
 $[$KK\,98$]$ 121&    12 05 24.5~~ $+$43 42 28 &    0.86  &    0.79 &  15.17      &  Sph{--}&L &   10.0 &    txt    \\
 Dw\,1214$+$2101 &    12 14 18.2~~ $+$21 01 08 &    0.30  &    0.67 &  20.1     &  Sph{--}&L &     7.0 &    mem   \\
 Dw\,1214$+$2945 &    12 14 26.6~~ $+$29 45 50 &    1.02  &    0.45 &  17.3     &  Tr{--}&N &     6.7 &    mem   \\
 Dw\,1215$+$2041 &    12 15 32.2~~ $+$20 41 00 &    0.40  &    0.71 &  18.7     &  Tr{--}&L &     7.0 &    mem  \\
 Dw\,1224$+$3938 &    12 24 34.6~~ $+$39 38 10 &    0.48  &    0.77 &  19.0     &  Tr{--}&L &     8.9 &    mem  \\
 Dw\,1229$+$4109 &    12 29 43.2~~ $+$41 09 43 &    0.29  &    0.60 &  19.3     &  Sph{--}&L &     8.9 &    mem  \\
 Dw\,1234$+$4116 &    12 34 38.2~~ $+$41 16 34 &    0.54  &    0.81 &  17.2     &  BCD{--}&H &     8.45&    NAM  \\
 KDG\,162        &    12 35 01.6~~ $+$58 23 08 &    1.04  &    0.69 &  18.5       &  Irr{--}&L &    10.0 &    txt  \\
 Dw\,1235$+$7050 &    12 35 59.5~~ $+$70 50 53 &    0.60  &    0.57 &  19.0     &  Irr{--}&L &     4.4 &    mem  \\
 Dw\,1237$+$3304 &    12 37 02.2~~ $+$33 04 59 &    0.50  &    0.60 &  17.5     &  Sph{--}&N &    10.90&    sbf  \\
 Dw\,1238$+$3337 &    12 38 18.0~~ $+$33 37 59 &    0.60  &    0.92 &  17.2     &  Tr{--}&N &    11.57&    sbf    \\
 AGC\,724993      &    12 38 30.0~~ $+$29 03 18 &    0.61  &    0.87 &  17.1       &  Im{--}&N &     9.20&    TF      \\
 Dw\,1239$+$2827 &    12 39 13.4~~ $+$28 27 14 &    0.55  &    0.59 &  18.7     &  Sph{--}&L &     8.9 &    mem     \\
 Dw\,1240$+$3037 &    12 40 39.8~~ $+$30 37 55 &    0.40  &    0.83 &  20.1     &  Sph{--}&VL &     7.35&    mem    \\
 Dw\,1241$+$4103 &    12 41 01.0~~ $+$41 03 11 &    0.33  &    0.76 &  19.1     &  Tr{--}&L &     7.7 &    mem     \\
 Dw\,1241$-$0427 &    12 41 22.3~~ $-$04 27 50 &    0.83  &    0.90 &  19.7     &   Sph{--}&VL&    10.1&     mem   \\
 KDG\,187        &    12 42 17.8~~ $+$03 28 08 &    1.35  &    0.93 &  15.6       &   dE{--}&N&     9.3&     mem   \\
 $[$KK\,98$]$ 162&    12 45 26.8~~ $+$18 18 01 &    1.12  &    0.64 &  17.7       &   Tr{--}&L&    10.0&     txt    \\
 Dw\,1245$+$6158 &    12 45 49.0~~ $+$61 58 08 &    0.81  &    0.49 &  18.4     &  Tr{--}&L &     5.6 &    mem    \\
 Dw\,1247$-$0824 &    12 47 25.0~~ $-$08 24 29 &    1.71  &    0.66 &  15.5     &  BCD{--}&N &     9.55&    mem    \\
 Dw\,1248$-$0915 &    12 48 38.4~~ $-$09 15 22 &    0.76  &    0.77 &  17.4     &  Sph{--}&L &     9.55&    mem   \\
 EVCC\,2232      &    12 50 04.7~~ $-$00 13 57 &    0.61  &    0.82 &  19.2     &  Sph{--}&L &     6.01&    NAM   \\
 Dw\,1252$-$0904 &    12 52 03.4~~ $-$09 04 26 &    0.34  &    0.70 &  19.5     &  Irr{--}&L &     9.55&    mem   \\
 AGC\,221126     &    12 56 18.0~~ $+$34 39 25 &    1.44  &    0.85 &  16.71    &  Im{--}&N &     9.5 &    bTF   \\
 $[$KK\,98$]$ 175&    12 59 01.0~~ $+$35 28 48 &    0.77  &    0.72 &  17.51    &  Irr{--}&N &     9.9 &    mem   \\
 Dw\,1311$+$4051 &    13 11 41.3~~ $+$40 51 47 &    0.37  &    0.61 &  19.1     &  Irr{--}&N &     9.0 &    mem    \\
 Dw\,1311$+$4317 &    13 11 45.6~~ $+$43 17 06 &    0.75  &    0.84 &  19.2     &  Sph{--}&VL &     9.0 &    mem    \\
 Dw\,1315$+$4304 &    13 15 00.5~~ $+$43 04 55 &    1.46  &    0.94 &  19.4     &  Sph{--}&VL &     9.0 &    mem     \\
 $[$KK\,98$]$ 206&    13 33 22.8~~ $+$49 06 07 &    1.03  &    0.66 &  14.87    &  Im{--}&N &     9.31&    TF      \\
 Dw\,2215$-$4528 &    22 15 25.4~~ $-$45 28 37 &    0.38  &    0.92 &  19.2     &  Sph{--}&L &    10.4 &    mem    \\
 Dw\,2216$-$4539 &    22 16 07.4~~ $-$45 39 11 &    1.08  &    0.71 &  17.5     &  Sph{--}&L &    10.4 &    mem     \\
 Dw\,2217$-$4633 &    22 17 32.4~~ $-$46 33 50 &   0.44   &   0.83  &  18.5     & Sph{--}&L  &    10.4  &   mem    \\
 6dF\,J22        &    22 18 48.7~~ $-$46 13 05 &   0.53   &   0.50  &  16.1     & Im{--}&N  &    10.4  &   mem    \\
 Dw\,2221$-$4608 &    22 21 03.4~~ $-$46 08 02 &   1.03   &   0.90  &  20.1     & Irr{--}&VL  &    10.4  &   mem     \\
 ESO\,289$-$020  &    22 21 11.5~~ $-$45 40 34 &   1.56   &   0.22  &  15.8     & Sd &     &    10.4  &   mem    \\
 Dw\,2221$-$4607 &    22 21 43.2~~ $-$46 07 01 &   0.35   &   0.53  &  20.4     & Tr{--}&VL  &    10.4  &   mem    \\
 Dw\,2227$-$4623 &    22 27 06.7~~ $-$46 23 10 &   0.51   &   0.83  &  16.8     & Tr{--}&N  &    10.4  &   mem   \\
\hline
%\end{tabular}
\end{longtable*}

Out of the 67 LV candidate dwarf galaxies that we found, 50
are not represented in the LEDA \citep{mak2014} and
 % %NED \url{(http://ipac.caltech.edu)}.
  NED\footnote{\url{http://ipac.caltech.edu}} databases.
 Some of the brightest galaxies from Table~\ref{Karachentsev_table2} %задокументированv
  are mentioned in the SDSS, GALEX \citep{bia2017} or
WISE\footnote{\url{http://wise2.ipac.caltech.edu/docs/release/allsky/}}
surveys,
% WISE ({\url{http://wise2.ipac.caltech.edu/docs/release/allsky/}}),
   but are not marked as possible nearby objects.

\section{REMARKS ON INDIVIDUAL CASES}
  {\it Dw0143$+$1338}. This very low surface brightness galaxy f
  alls in our survey region around NGC\,628. However, it is
  located at 13\arcmin\ to the east of the peculiar galaxy
NGC\,660, the distance to which is $12.3$~Mpc (NAM), and is
probably a satellite of NGC\,660 rather than NGC\,628.

  {\it Dw\,0149$+$3237}. A very low surface brightness galaxy marked by \citet{ann2020}.

  {\it Dw\,0910$+$7326}.  The galaxy is partially resolved into stars. It is extraordinary that it was not detected earlier.

  {\it UGC\,6451}. In contact with a distant galaxy UGC\,6450, whose radial velocity is $14\,500$~km\,s$^{-1}$.

  {\it Dw\,1159$+$5554}. The object is located in the vicinity of an S0-galaxy NGC\,3390. Has an unusual form of a smooth
``bean'' without signs of star formation. The nature of this
object is mysterious.

  {\it Dw\,1214$+$2945}. This dwarf galaxy has a radial velocity of
  \mbox{$V_{\rm LG}=395$~km\,s$^{-1}$} (SDSS). As is the case with
  its neighbor spiral NGC\,4136, it is located in the region of the
  Coma\,I group, where galaxies at distances of about $15$~Mpc have
  high negative peculiar velocities.

  {\it Dw\,1234$+$4116}. A blue
  dwarf galaxy with a radial velocity of \mbox{$V_{\rm LG}=640$~km\,s$^{-1}$}
(SDSS), probable satellite of  NGC\,4618 with \mbox{$V_{\rm
LG}=576$~km\,s$^{-1}$}.

  {\it Dw\,1237$+$3304} and {\it Dw\,1238+3337}. According to the
  distance measurements for these galaxies in \citet{car2022}, both
  dwarf galaxies are located behind the NGC\,4631 group.

 {\it AGC\,724993}. The radial velocity of this galaxy,
 \mbox{$V_{\rm LG}=736$~km\,s$^{-1}$} \citep{hay2018}, and its
TF-distance, $9.2$~Mpc, indicate that it belongs to the satellites
of NGC\,4559 (\mbox{$V_{\rm LG}=576$~km\,s$^{-1}$} and
TRGB-distance $8.91$~Mpc).

\begin{figure} \vspace{2mm}
%\setcaptionmargin{5mm} \onelinecaptionsfalse \captionstyle{normal}
\includegraphics[width=8.3cm, bb= 5 10 3034 4425,clip]{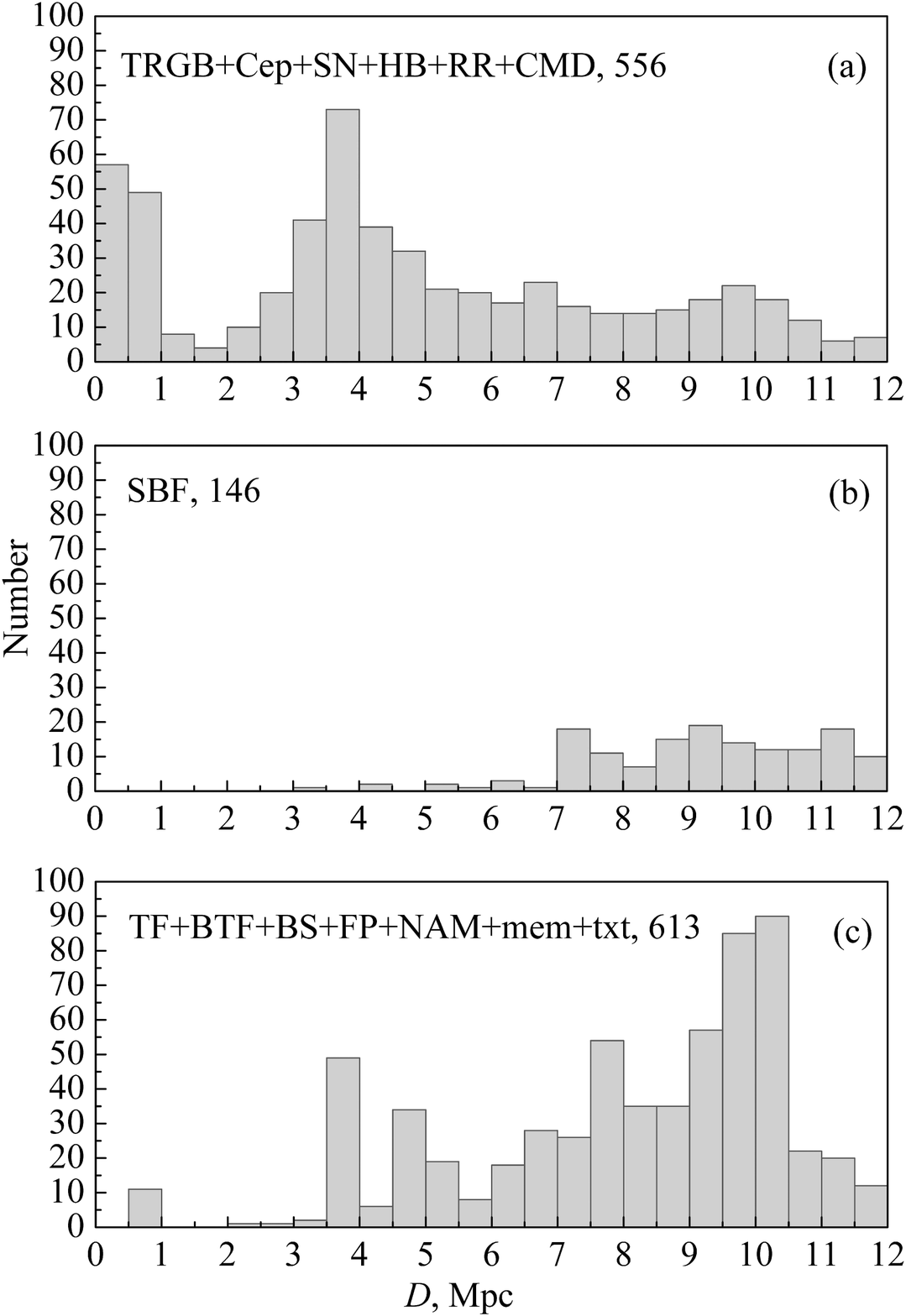}
   \caption{Distribution of Local Volume galaxies by distance $D$,
   estimated using methods with various accuracies: approximately
   5\%,15\% and 25\%---panels (a), (b) and (c), correspondingly.}
  \label{Karachentsev_figure1}
  \end{figure}
  {\it KDG\,187}. This is VCC\,1917 with radial velocity
  \mbox{$V_h=731$~km\,s$^{-1}$}, a possible Virgo cluster member.

  {\it Dw1247$-$0824}. An H\,I-signal with a radial velocity of
  \mbox{$V_h=1215$~km\,s$^{-1}$} is noticeable in HIPASS at the
  location of this BCD dwarf.

  {\it EVCC\,2232}. A possible Virgo cluster member with a radial
  velocity of \mbox{$V_{\rm LG}=608$~km\,s$^{-1}$}
or a dwarf in front of the cluster.

  {\it AGC\,221126=UGCA\,309}. A gas-rich dwarf with radial velocity
  \mbox{$V_{\rm LG}=747$~km\,s$^{-1}$} in the vicinity of NGC\,4861.

  {\it [KK\,98]175=AGC\,223250}. Another possible NGC\,4861 satellite
  with a radial velocity of \linebreak
\mbox{$V_{\rm LG}=725$~km\,s$^{-1}$.} It had earlier been assigned
an erroneous radial velocity of \mbox{$V_{\rm
LG}=1205$~km\,s$^{-1}$.}

  {\it Dw1315$+$4304}. A spheroidal dwarf galaxy of very low surface
  brightness, separated from NGC\,5055 by a projected distance $175$~kpc.

  {\it [KK\,98]206}. This is the blue dwarf galaxy SBS\,1331+493 with radial velocity
\mbox{$V_h=594$~km\,s$^{-1}$}, a possible distance satellite of
M\,51.

  {\it 6dF\,J\,22 and ESO\,289$-$020}. Satellites of an Scd-galaxy IC\,5201.
  Their heliocentric radial velocities, $996$~km\,s$^{-1}$ and
  $917$~km\,s$^{-1}$, were measured by \citet{kle2019}.

  \section{DISCUSSION}
In order to confirm the status of the Local Volume members, one
needs radial velocity and distance measurements for the presented
candidates. The most promising method would be to determine the
distances by TRGB luminosity with HST or by surface brightness
fluctuations using images obtained with large ground-based
telescopes with subarcsecond seeing. The more ``test particles''
(dwarf galaxies) with known peculiar velocities are contained in
the LV, the more accurately can the 3D-contour of the dark matter
in the Local Volume be reconstructed. Such a task is currently
unachievable in the more distant volumes of the Universe.

 \begin{figure*}
% \setcaptionmargin{5mm} 
%\onelinecaptionsfalse
%\captionstyle{normal}\vspace{3mm}
\includegraphics[width=10.0cm, bb= 72 130 2640 5700]{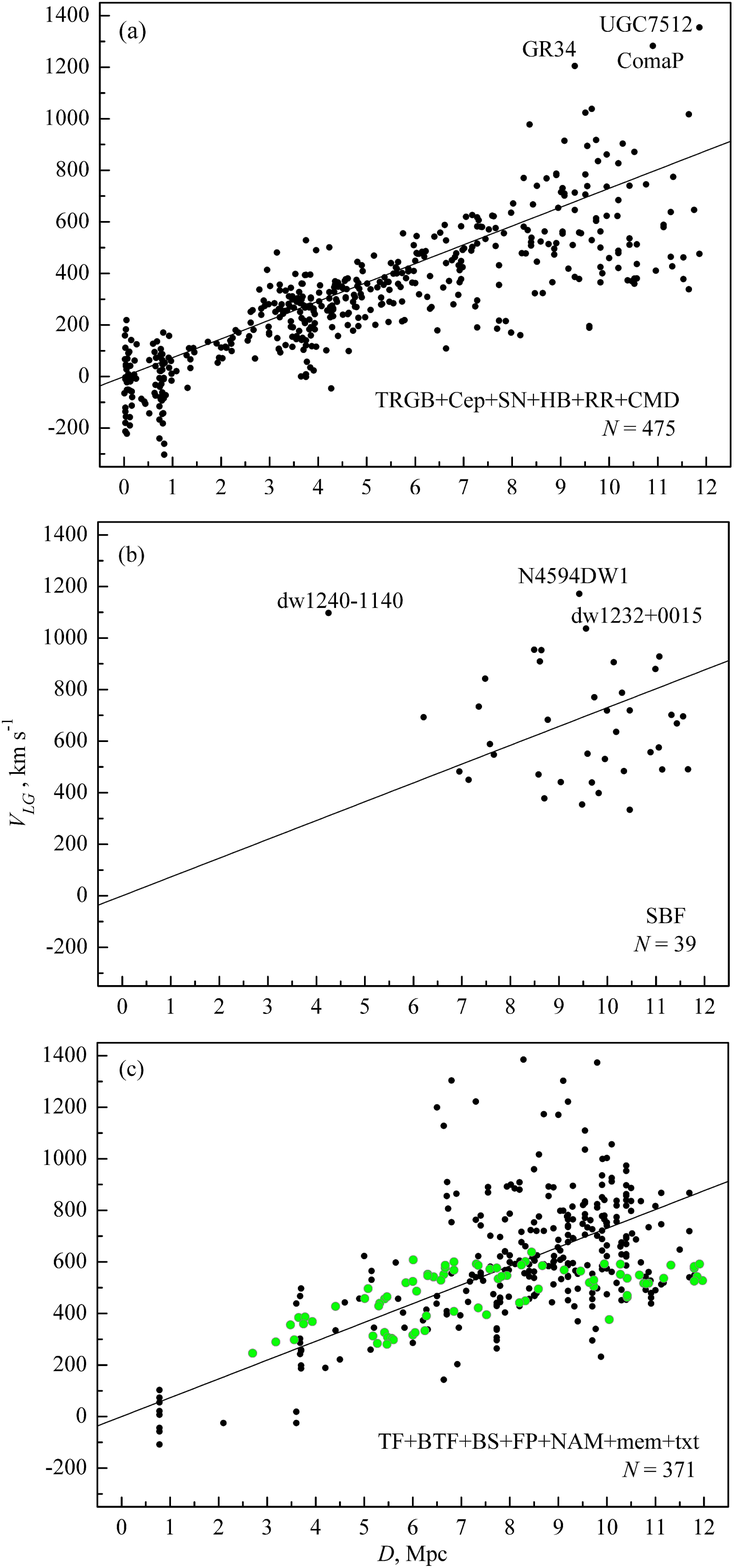}
  \caption{Distribution of the Local Volume galaxies by radial
velocities relative to the Local Group centroid and by distance
from the Milky Way. Panels~(a), (b) and (c) show galaxies with high,
medium and low accuracy distance estimates, correspondingly.
 The straight line corresponds to the Hubble parameter
 $73$~km\,s$^{-1}$\,Mpc$^{-1}$. In panel~(c), galaxies with
 kinematic distance estimates (NAM) are shown by the green circles.}
  \label{Karachentsev_figure2}
  \end{figure*}

  The galaxy distribution in the LV by distance from the Milky Way is
  presented in Fig.~\ref{Karachentsev_figure1}.
  Fig.~\ref{Karachentsev_figure1}a shows  499~galaxies with $D$
  estimated by the TRGB method, the accuracy of which is about 5\%.
  To these galaxies we added another 57 with distances measured by
  the supernovae luminosity (SN), Cepheids (cep), RR Lyrae type
  stars (RR), horizontal branch (HB) and CMD galaxies (CMD).
Fig.~\ref{Karachentsev_figure1}b corresponds to the subsample of
galaxies with distances estimated by the surface brightness
fluctuations \mbox{($N = 146$)}. The $D$ measurement accuracy for
them is about 15\%. Fig.~\ref{Karachentsev_figure1}c presents the
remaining LV galaxies with distances estimated by the
 Tully--Fisher method (TF, BTF, \mbox{$N = 174$}),
by the radial velocity in the NAM model \mbox{($N = 81$),} by the
assumed membership in groups (mem, \mbox{$N = 329$}), by the
brightest stars (BS, \mbox{$N = 10$}), and also by the general
texture of the galaxies (txt, \mbox{$N = 19$}). The averaged
accuracy of the distances estimated using these methods can be
adopted as approximately equal to 25\%.
Fig.~\ref{Karachentsev_figure1} does not show the 106~galaxies
whose distances turned out to be greater than 12~Mpc.

Evidently, galaxies with accurately measured distances dominate in
a 6~Mpc radius sphere. Near the far boundary of the Local Volume,
most galaxies have distances measured with a low accuracy. On the
whole, the relative number of LV galaxies with reliably measured
distances does not yet exceed 40\%. Curiously, in 2001, by the
start of systematic TRGB distance measurements with the HST, the
number of known LV galaxies was about~450. Over the past 22~years
this method was used to measure TRGB-distances for approximately
500~LV galaxies, but the LV population itself became three times
larger over this time. (The wolf will follow the rabbit, but will
probably never catch it).

 \begin{figure*}
% \setcaptionmargin{5mm}
%\onelinecaptionsfalse \captionstyle{normal}
 %\includegraphics[width=12.5cm, bb= 10 15 295 205,clip]{Karachentsev_fig3}
 \includegraphics[width=12.5cm, bb= 10 15 295 205,clip]{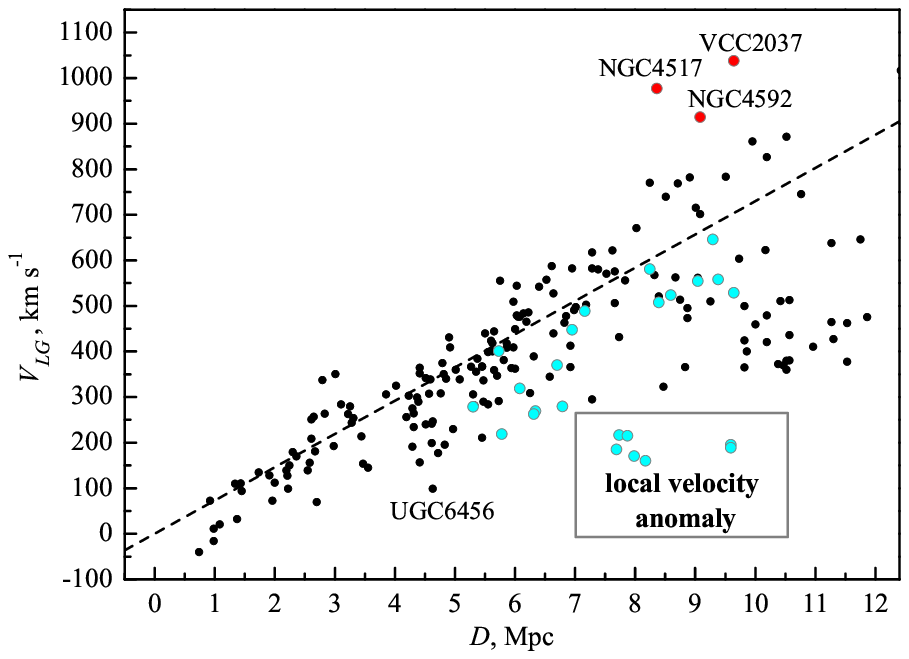}
 \caption{Hubble ``velocity--distance'' relation, similar to
 Fig.~\ref{Karachentsev_figure2}, but for the general field galaxies
with negative tidal indices.  Galaxies in a 13\degr\ radius cone around
the center of the Virgo cluster are shown by
the red circles, whereas galaxies at high negative supergalactic
latitudes are shown by cyan circles. The straight line corresponds
to the unperturbed Hubble flow with the parameter
$73$~km\,s$^{-1}$\,Mpc$^{-1}$. }
  \label{Karachentsev_figure3}
  \end{figure*}

Currently, radial velocities are measured for 893 galaxies with
distances within $12$~Mpc. Fig.~\ref{Karachentsev_figure2} shows
the behavior of the Hubble flow in the Local Volume for three
types of galaxies shown in Fig.~\ref{Karachentsev_figure1}. A
trend can be seen: the higher the distance measurement accuracy,
the smaller the Hubble diagram scatter.
In Fig.~\ref{Karachentsev_figure2}a, the typical accuracy of the
galaxy distance estimates is only \mbox{($0.3$--$0.6$)~Mpc}, and
therefore, their deviation from the ideal Hubble flow is due to
their real peculiar velocities.
In Fig.~\ref{Karachentsev_figure2}b, the Hubble diagram looks
expressionless. Due to the absence of emission lines in the radio
and optical spectra for the dwarf spheroidal galaxies, only a
quarter of them have measured radial velocities. In some cases
\mbox{({dw}\,1240$-$1140)}, distances measured by the surface
brightness fluctuations may contain significant errors.

In Fig.~\ref{Karachentsev_figure2}c, the main mass comprises
galaxies with distances estimated by the \mbox{Tully--Fisher}
relation between the internal motions amplitude and total galaxy
luminosity. Additionally, the diagram shows 85~galaxies with
kinematic NAM model distance estimates (green circles). The
scatter of these galaxies with respect to the straight line
characterizes the role of systematic non-Hubble flows in the Local
Volume. Fig.~\ref{Karachentsev_figure2}c does not show seven dwarf
galaxies: VCC\,114, VCC\,169, KDG\,104, UGC\,7642, VCC\,1675,
IC\,3591 and VCC\,1713. They are all located near the direction
towards the Virgo cluster center, they have TF-distances between
$7$~and $10$~Mpc and radial velocities in the interval from~$1400$
to $2100$~km\,s$^{-1}$, taking part in accelerated motion towards
the cluster center. The same effect also influences the galaxies
UGC\,7512, Coma\,P and GR\,34 with TRGB-distances marked in
Fig.~\ref{Karachentsev_figure2}a.
 \begin{figure*}
%\setcaptionmargin{5mm} \onelinecaptionsfalse \captionstyle{normal}
\includegraphics[scale=1.25,bb= 15 15 270 210,clip]{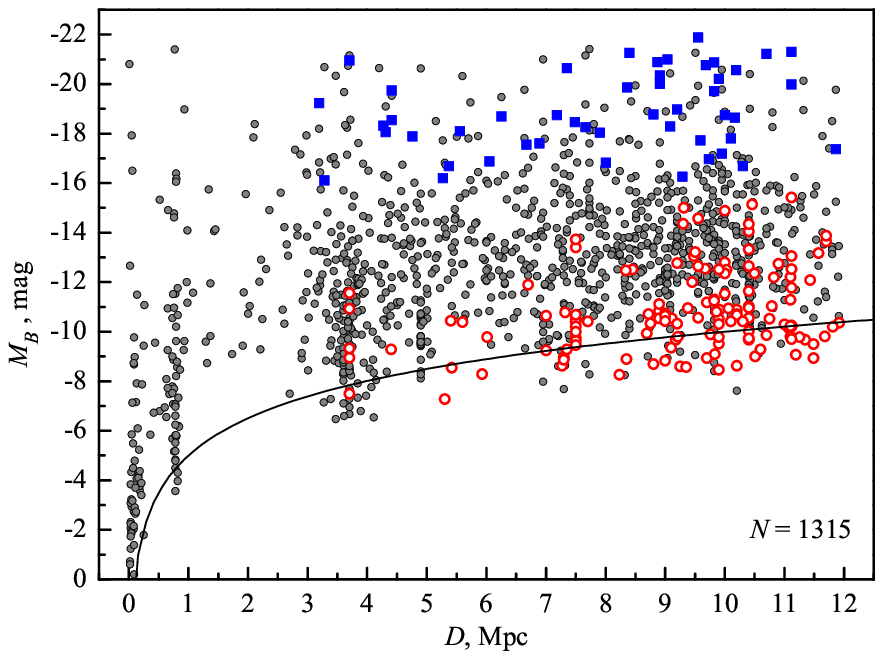}
  \caption{Distribution of the Local Volume galaxies by absolute
  $B$-magnitude and distance from the Milky Way. The blue squares
  show the galaxies around which we searched for
new satellites.  The open red circles show the new Local Volume
candidates, recently found in the DESI Legacy Imaging Surveys.
    The curve corresponds to the total apparent magnitude of
    the galaxies $B_t = 20\fm0$.}
  \label{Karachentsev_figure4}
  \end{figure*}

 If we exclude the members of the known nearby groups with their
virial motions from Fig.~\ref{Karachentsev_figure2}a, the Hubble
diagram for the remaining general field galaxies (with negative
``tidal index'' $\Theta_1$ in the UNGC catalog) will take the form
presented in Fig.~\ref{Karachentsev_figure3}. At small distances
\mbox{$D<2.5$~Mpc}, the Hubble flow looks ``cold'' with a
characteristic peculiar velocity of \mbox{$\sigma_v\sim
30$~km\,s$^{-1}$.} The close galaxies at a distance of the order
of $1$~Mpc have decreased radial velocities due to their
deceleration by the total mass of the Local Group. Large scale
flows begin to manifest themselves at larger distances. The red
circles in Fig.~\ref{Karachentsev_figure3} show three galaxies in
the direction of the Virgo cluster within a 13\degr\ radius from
its center. The cyan circles show galaxies at supergalactic
latitudes \mbox{$SGB<-45\degr$}. Seven of them, with velocities of
the order of $200$~km\,s$^{-1}$ and distances
\mbox{$8$--$9$~Mpc}, are located in the zone of the so-called
``local velocity anomaly''. The deviations of the first category
of galaxies from the unperturbed Hubble flow with the parameter
\mbox{$H_0=73$~km\,s$^{-1}$\,Mpc$^{-1}$} is due to the falling of
these galaxies towards the closest massive attractor Virgo. The
deviations of the second category of galaxies are caused by their
systematic motion from the center of the expanding Local Void. An
analysis of the local peculiar velocity field allows one to
estimate the total mass of the Virgo cluster,
\mbox{$M_T=8\times10^{14}~M_{\odot}$} \citep{kar2014}, as well as
the size and density contrast of the Local Void \citep{nas2011}.
Generally, the combination of isolated nearby galaxies follows the
Hubble relation with a slope of not $73$, but
$60$~km\,s$^{-1}$\,Mpc$^{-1}$. This is probably due to an excess
of objects at high supergalactic latitudes among the nearby
isolated galaxies.

The distribution of LV galaxies by absolute $B$-magnitudes and
distances from the MW is shown in Fig.~\ref{Karachentsev_figure4}.
The blue squares show the galaxies around which we searched for
new satellites. The red empty circles mark the galaxies added to
the LV sample based on our data and the data of \citet{car2022} in
the process of searching in the {DESI Legacy Imaging Surveys}. A
deficit of ultra-dwarf galaxies with absolute magnitudes fainter
than $-9^{\rm m}$ remains noticeable near the far boundary of the
Local Volume. This gap will obviously be filled in deeper sky
surveys: Euclid \citep{lau2011},  Roman Space Telescope \citep{spe2015}
and LSST, planned in the nearest years to come.

\begin{acknowledgements}
The authors are grateful to D.~I.~Makarov, who has repeated the
computations of the tidal indices for the Local Volume galaxies
with allowance for new data. This work has made use of the {DESI
Legacy Imaging Surveys} data.
\end{acknowledgements}

\section*{FUNDING}
The work on updating the galaxy catalog was carried out within
grant \textnumero~075-15-2022-262 \linebreak (13.{MNPMU}.21.0003) of the
Ministry of Science and Higher Education of the Russian
Federation.

\section*{CONFLICT OF INTEREST}
The authors declare no conflict of interest.

\begin{flushright}
{\it Translated by L. Chmyreva}
\end{flushright}
%\selectlanguage{russian}

%\newpage
\onecolumngrid
\clearpage
\section*{APPENDIX}
\begin{center}
 {Images of candidate satellites for nearby massive galaxies, taken from DESI Legacy Imaging Surveys.
The map size is $2\arcmin\times 2\arcmin$. North is at the top,
east is to the left.}
\end{center}

%\begin{figure*}
%\begin{minipage}[c]{17.0cm}
\begin{center}
\vspace*{0.1cm} %\setcounter{figure}{-1}
%\hspace*{0.3cm}
 \includegraphics[scale=0.4]{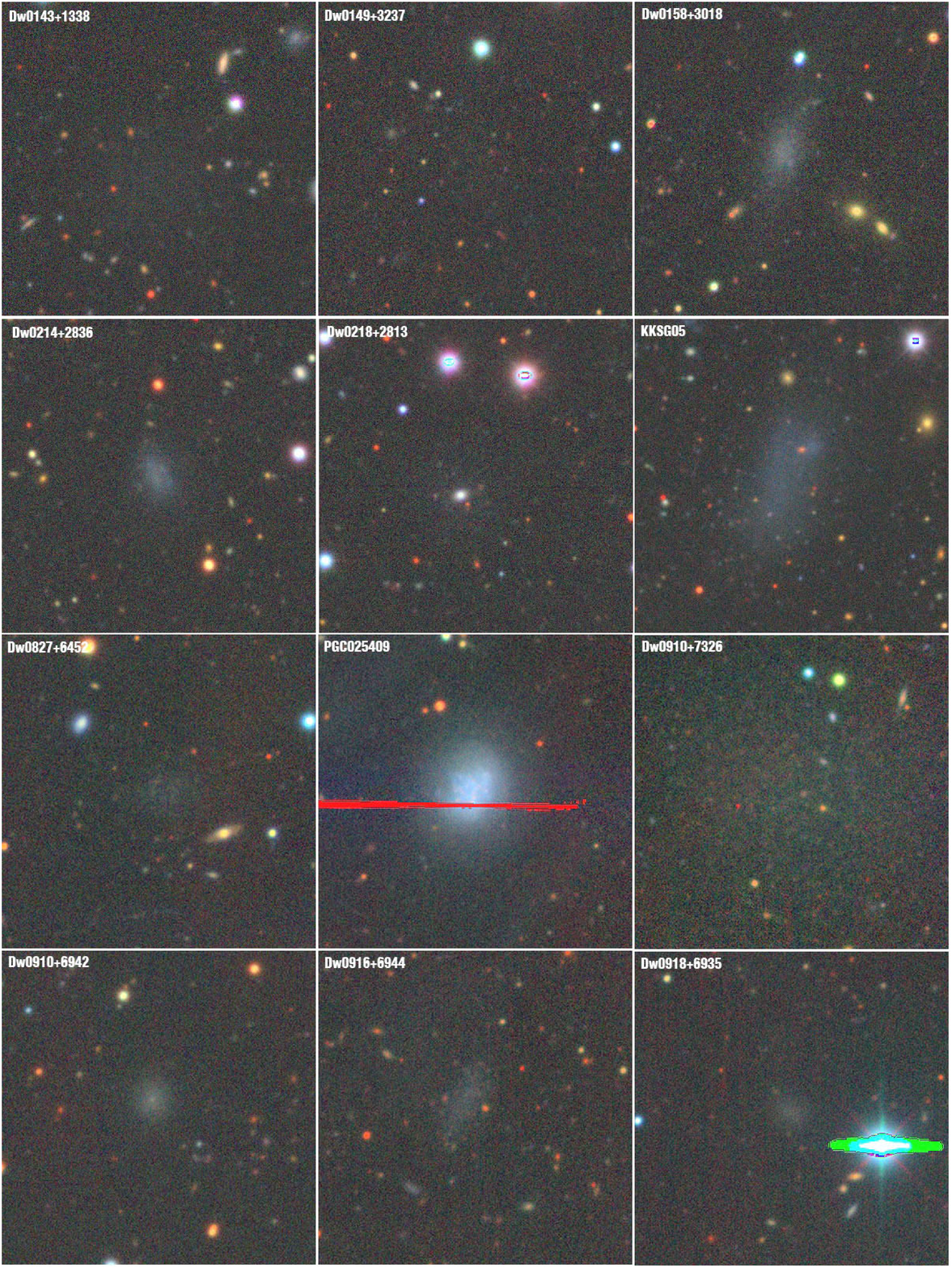}
 %\label{figure1}
%\end{center}
%\end{minipage}
%\end{figure*}
\newpage
%  \setcounter{figure}{-1}
%  \begin{figure*}
%  \begin{center}
 \vspace*{0.1cm} %\setcounter{figure}{-1}
\hspace*{0.3cm}  \includegraphics[scale=0.4]{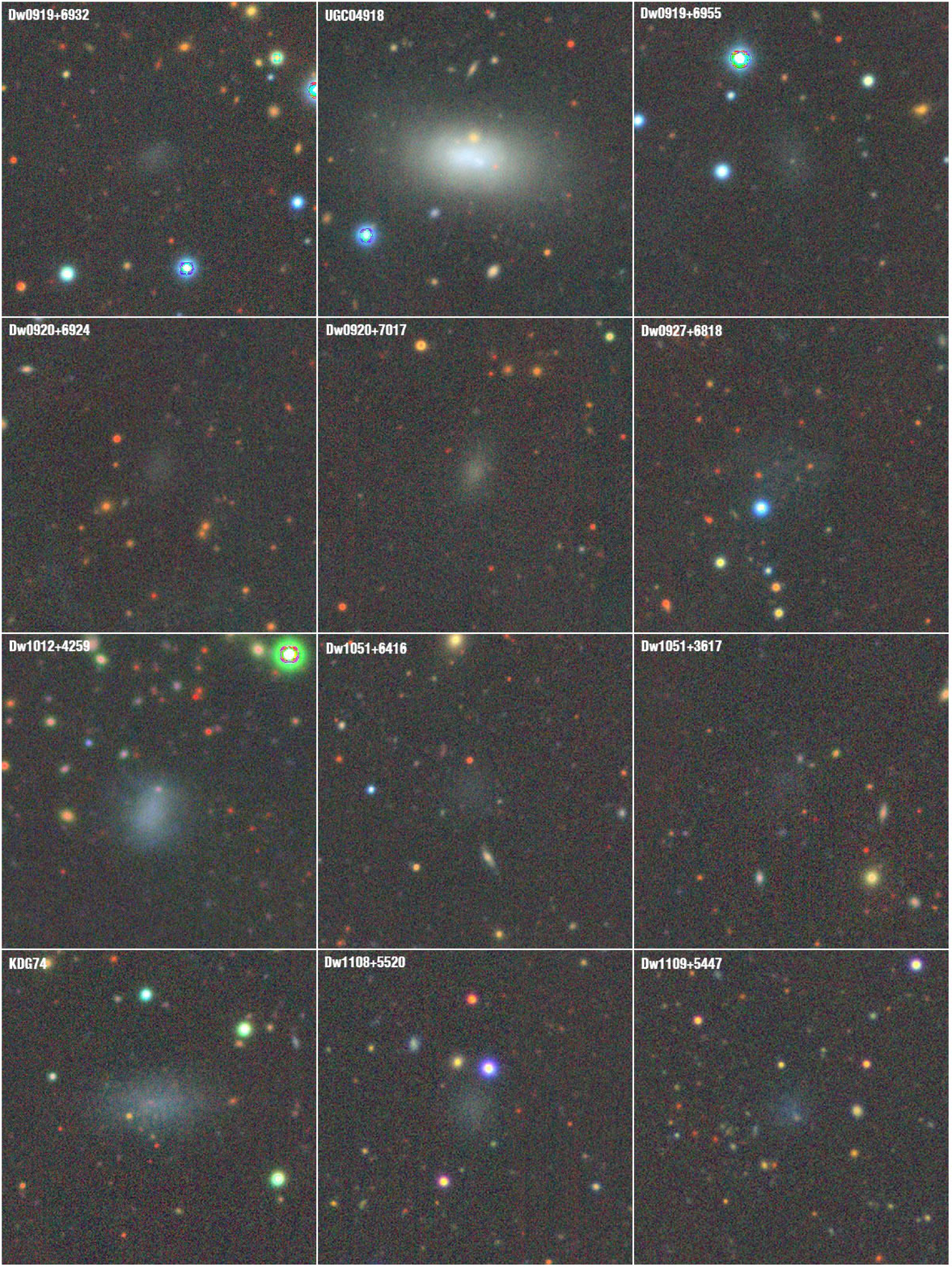}
   %\label{figure1}
%   \end{center}
%  \end{figure*}

%  \setcounter{figure}{-1}
%  \begin{figure*}
%  \begin{center}
% \vspace*{0.1cm} %\setcounter{figure}{-1}
%\hspace*{0.3cm}  
\includegraphics[scale=0.4]{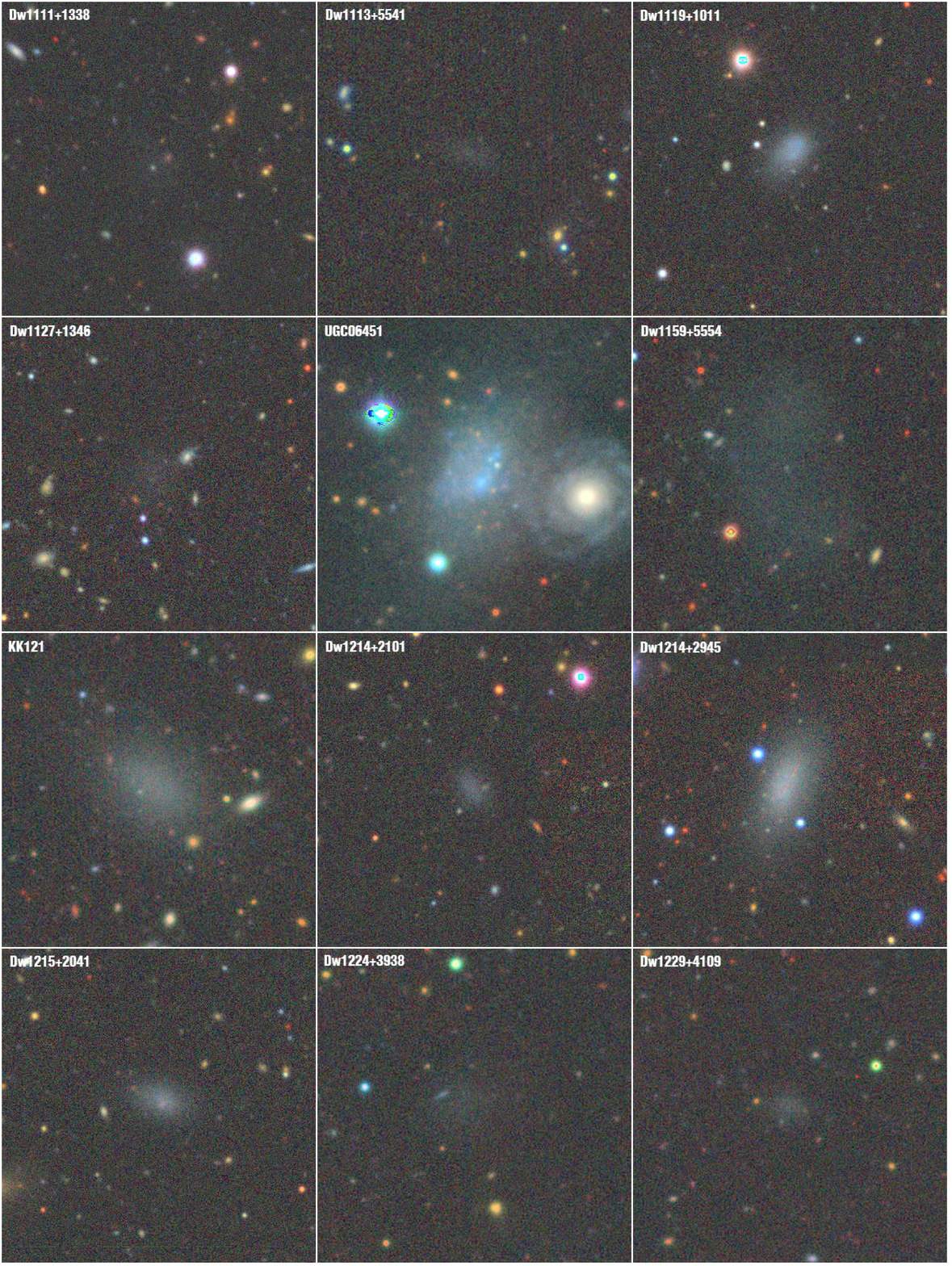}
     %\label{figure1}
%     \end{center}
%  \end{figure*}
\newpage
%  \setcounter{figure}{-1}
%  \begin{figure*}
%  \begin{center}
 \vspace*{0.1cm} %\setcounter{figure}{-1}
\hspace*{0.3cm}  \includegraphics[scale=0.4]{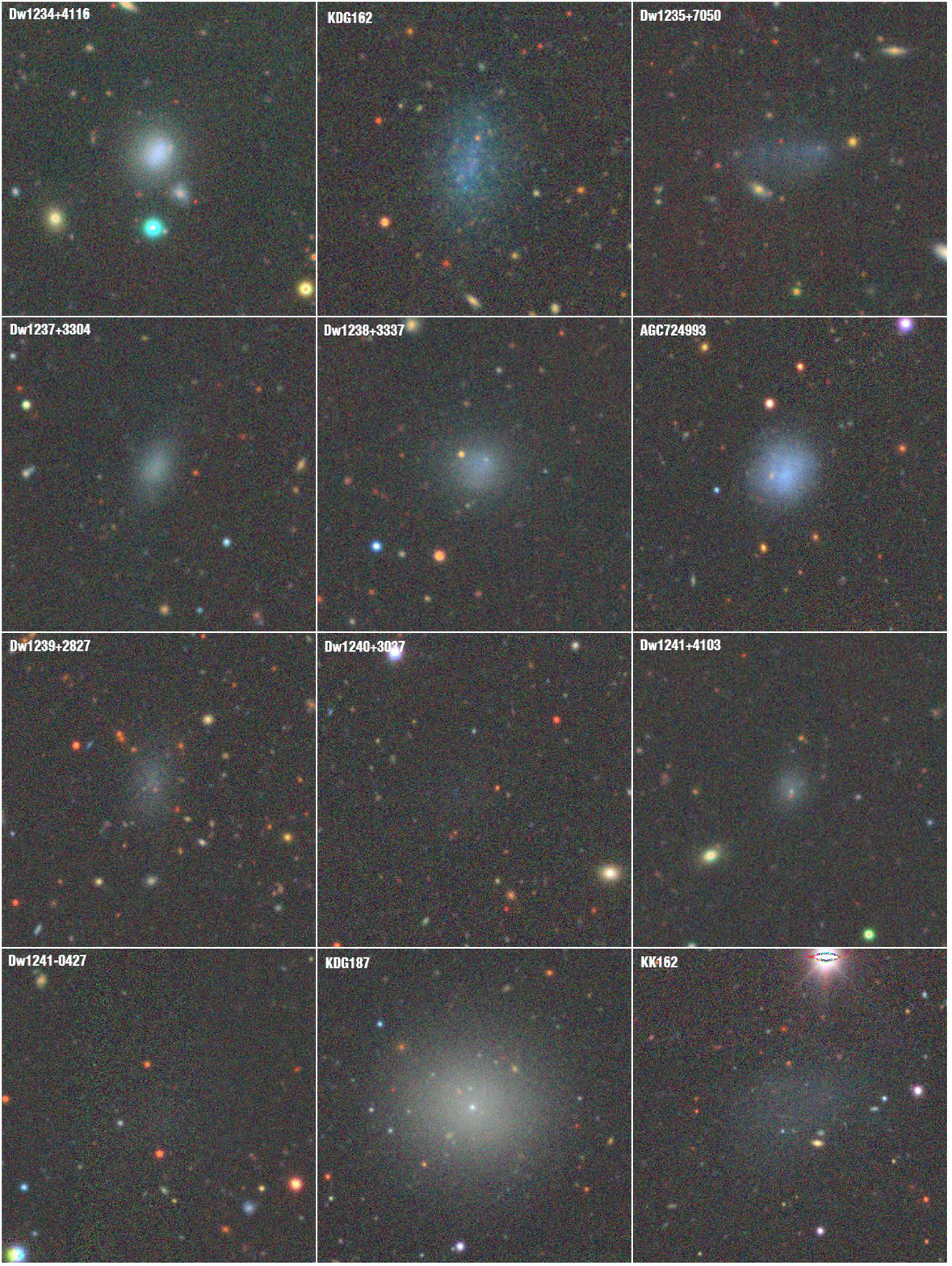}
     %\label{figure1}
%     \end{center}
%  \end{figure*}
\newpage
%  \setcounter{figure}{-1}
%   \begin{figure*}
%   \begin{center}
\vspace*{0.1cm} %\setcounter{figure}{-1}
\hspace*{0.3cm}   \includegraphics[scale=0.4]{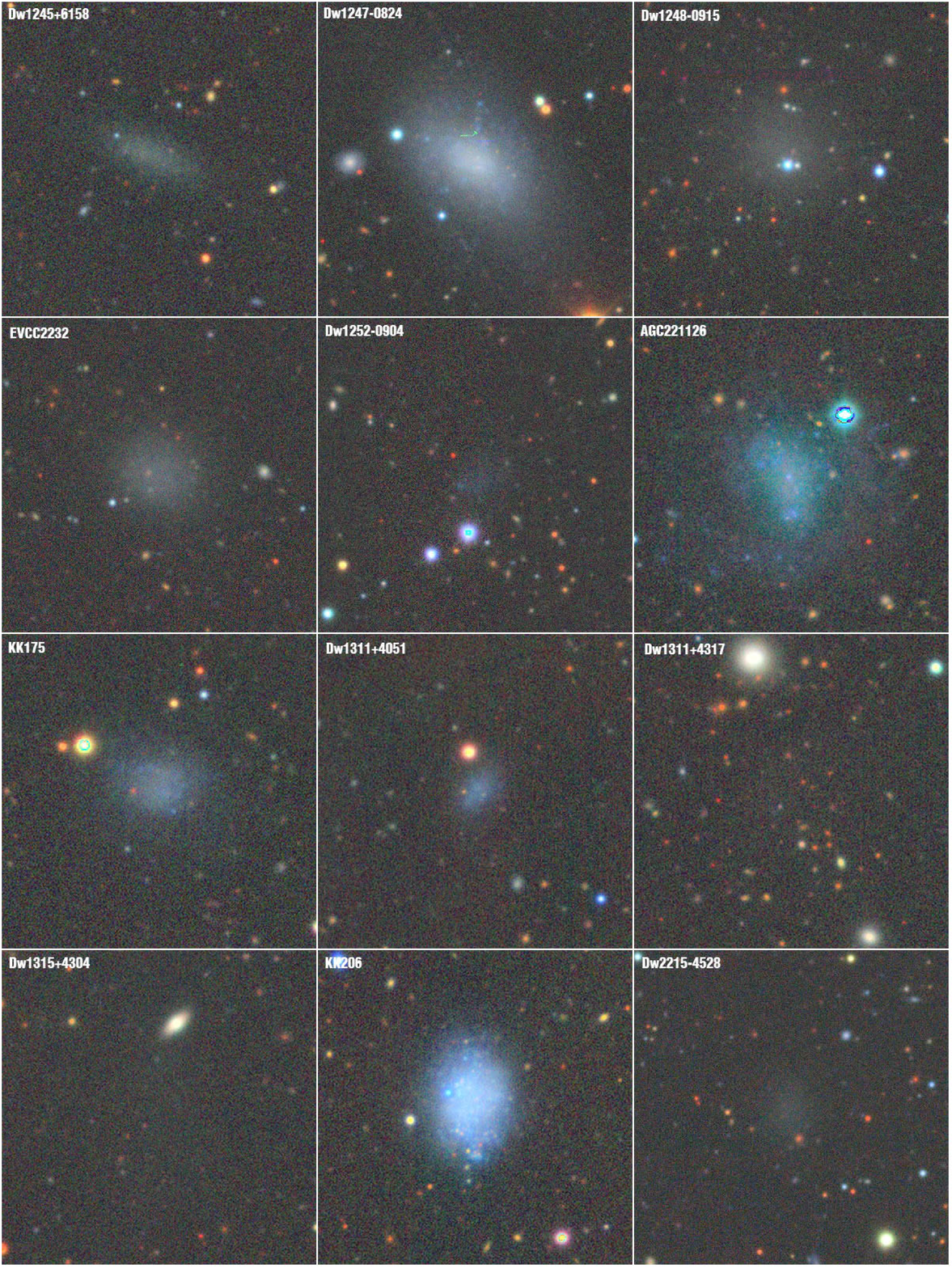}
    %\label{figure1}
%    \end{center}
%  \end{figure*}
\newpage
%  \setcounter{figure}{-1}
%  \begin{figure*}
%  \begin{center}
 %\setcounter{figure}{-1}
\vspace*{0.1cm}
\hspace*{0.3cm}
\includegraphics[scale=0.4]{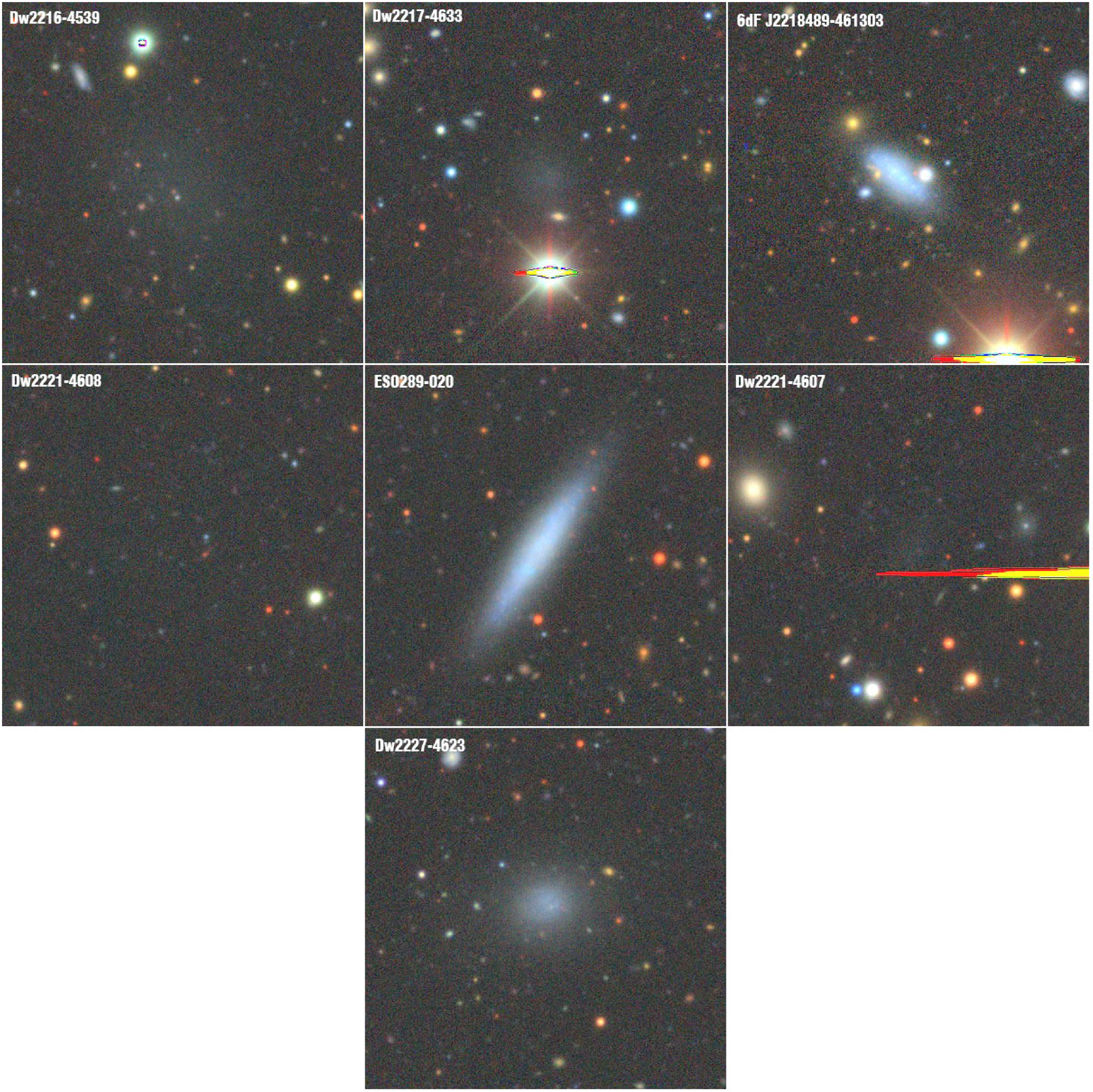}
   %\label{figure1}
   \end{center}
%  \end{figure*}
%\onecolumngrid

%\endinput

  \end{document}